# Singularity as a diagnostic for secondary eyewall occurrence in tropical cyclones

Ping Lu,[a] Long Yang,[b] Jishi Zhang,[a] Danyang Wang,[a] Yanluan Lin [a]

[a] *Department of Earth System Science, Tsinghua University, Beijing, China*

[b] *School of Geography and Ocean Science, Nanjing University, Nanjing, China*

*Corresponding authors*:

Ping Lu, luping@mail.tsinghua.edu.cn

Yanluan Lin, yanluan@mail.tsinghua.edu.cn

ABSTRACT

Secondary eyewalls occur in 70% of major tropical cyclones (TCs), and are associated with rapid changes in storm intensity and rapid broadening of strong winds. While mechanisms of secondary eyewall formation have been investigated from various perspectives, the explicit conditions on which secondary eyewalls occur in TCs remain veiled, leaving substantial uncertainties in TC intensity forecast, especially for the most extreme events. In this study, we present a simple diagnostic, in form of a singularity, for secondary eyewall occurrence in TCs. The diagnostic is solely dependent on three basic storm characteristics (the maximum wind speed, the radius of maximum wind, and the latitude) and shown to compare well with satellite observations. It provides a valuable tool to improve the understanding, modeling and risk assessment of secondary eyewall storms.

## 1. Introduction

Tropical cyclones (TCs), also known as hurricanes or typhoons, are commonly characterized by a tranquil low pressure center, i.e. storm eye, and a ring of intense convection called the eyewall (the term is a direct representation of the ~16 km's tall cloud wall that one can visually see from the storm eye). An eye and one ring of eyewall constitute the typical structure of TC. Observations have long shown, however, that for some TCs there exists a secondary (or even tertiary) eyewall outside the primary eyewall (Willoughby et al. 1982; Black and Willoughby 1992), with the moat region between them, a nearly echo-free annulus on radar, taking on the characteristics of the eye (Houze et al. 2007). The rise of secondary eyewall is usually accompanied by the weakening of primary eyewall, which eventually dissipates while the secondary eyewall contracts and takes over, and such a process is called eyewall replacement. Eyewall replacements can last a few hours to more than a day (vary significantly among storms), during which storms undergo large oscillations in intensity and size, and is regarded as a 'key process in hurricane intensity change' (Houze et al. 2007). Storms with eyewall replacements can have serious consequences, especially the rapid changes in intensify (e.g. Hurricane Andrew 1992, Hurricane Irma 2017) and rapid broadening of strong winds (e.g. Hurricane Katrina 2005) just prior to landfall.

Secondary eyewalls may not be captured by visible or infrared images because of the shielding from cirrus canopy and outward-slanting primary eyewall. It was not until 2004, when long-term passive microwave data was analyzed, that we learned the percentage of secondary eyewalls is 'far higher than previously thought'(Hawkins and Helveston 2004; Hawkins et al. 2006). About 70% of major hurricanes (Saffir-Simpson Hurricane Scale, SSHS category 3-5, wind speed > 47 m s$^{-1}$) show secondary eyewalls (Hawkins and Helveston 2004; Hawkins et al. 2006; Kossin and Sitkowski 2009; Kuo et al. 2009). Comparing with single-eyewall storms, storms with secondary eyewalls are found to be associated with stronger wind speed, smaller eye diameter, colder infrared brightness temperatures, higher sea surface temperatures, weaker environmental wind shear, and lower latitudes (Kossin and Sitkowski 2009; Hence and Houze 2012; Yang et al. 2013). Unlike single-eyewall storms, the weakening of the maximum wind speed in secondary-eyewall storms typically occurs in an environment that is not indicative of weakening (Kossin and DeMaria 2016) and is accompanied with maintaining or increasing convective activity (Yang et al. 2013).

Mechanisms of secondary eyewall formation have been investigated from various perspectives, from the ambient environment (e.g. humidity (Hill and Lackmann 2009; Ge 2015), beta shear(Fang and Zhang 2012), storm interaction with midlatitude jet (Dai et al. 2017), upper-level trough (Nong and Emanuel 2003; Molinari and Vollaro 1990), nearby vortices(Kuo et al. 2004, 2008)), to the internal dynamics of the storm (e.g. vortex Rossby waves-mean flow interaction (Montgomery and Kallenbach 1997; Terwey and Montgomery 2008), potential vorticity in rainbands (May and Holland 1999; Judt and Chen 2010), supergradient wind and unbalanced boundary layer response (Bell et al. 2012; Huang et al. 2012; Abarca and Montgomery 2013, 2014), positive feedback among radial vorticity gradient, frictional convergence and moist convection (Kepert 2013), wind-induced surface heat exchange(Nong and Emanuel 2003; Cheng and Wu 2018), outer-core latent heating (Bell et al. 2012; Rozoff et al. 2012; Wang 2009), timescale of filamentation vs. convection (Rozoff et al. 2006) , ice-phase microphysics (Zhou and Wang 2011)). Although a number of conditions and processes have been found directly related and several hypotheses proposed, a consolidated theory that can fully grasp the critical dynamics of secondary eyewall formation is still on its way.

In our continuing efforts to develop a simple physics-based TC rainfall model (Emanuel 2017; Lu et al. 2018; Zhu et al. 2013; Feldmann et al. 2019; Xi et al. 2020; Zhu et al. 2021; Gori et al. 2022) for risk assessment purpose, we find that a singularity arises in computing boundary layer convergence for the most intense and compact storms in low latitudes. The emergence of this singularity directly results in a nonphysical break of computed radially inward flow, very much resembling storms with secondary eyewalls. After careful evaluation with satellite-observed secondary eyewall cases, we propose the emergence of such singularity as a simple diagnostic for secondary eyewall occurrence.

The proposed diagnostic is solely dependent on three parameters, the maximum wind speed ($V_m$), the radius of maximum wind ($R_m$), and the Coriolis parameter ($f$, computed from latitude of storm center) – three basic storm characteristics that are routinely recorded in TC observations. We are surprised that the information of a storm undergoing eyewall replacement, previously recognized only by satellite observations (e.g. Kossin and Sitkowski 2009; Kuo et al. 2009) or flight-level aircraft observations (e.g. Sitkowski et al. 2011, 2012), is folded in just three numbers.

The diagnostic is detailed in Section 2, evaluated in Section 3, discussed in Section 4.

## 2. Mathematical expression of the singularity

In the frictional inflow layer in a circular vortex, we assume the principal balance is

$$u \frac{\partial M}{\partial r} \simeq -r \frac{\partial \tau_\theta}{\partial z}, \qquad (1)$$

between radial advection of angular momentum and frictional torque acting on the azimuthal velocity (Ooyama 1969; Lu et al. 2018), where $u$ is the radial velocity, $r$ is the radius from the storm center, $M$ is the absolute angular momentum per unit mass [$M = rV + 0.5fr^2$, where $V$ is the azimuthal wind speed and $f$ is the Coriolis parameter], and $\tau_\theta$ is the azimuthal turbulent stress.

Equation 1 is a simple description of how the azimuthal velocity $V$ (which is used to compute $M$) and radial velocity $u$ are connected. Since $V$ is an order of magnitude larger than $u$, and usually the focus of observations, $V$ is commonly used as a known input to deduce $u$ in TC boundary layer and rainfall modeling. Since the most common observations of $V$ is the maximum wind speed $V_m$ and radius of maximum wind $R_m$, one common practice, especially in TC risk assessment, is to construct the horizontal distribution of $V$ given $V_m$ and $R_m$ using parametric wind models (as in Fig. 1A).

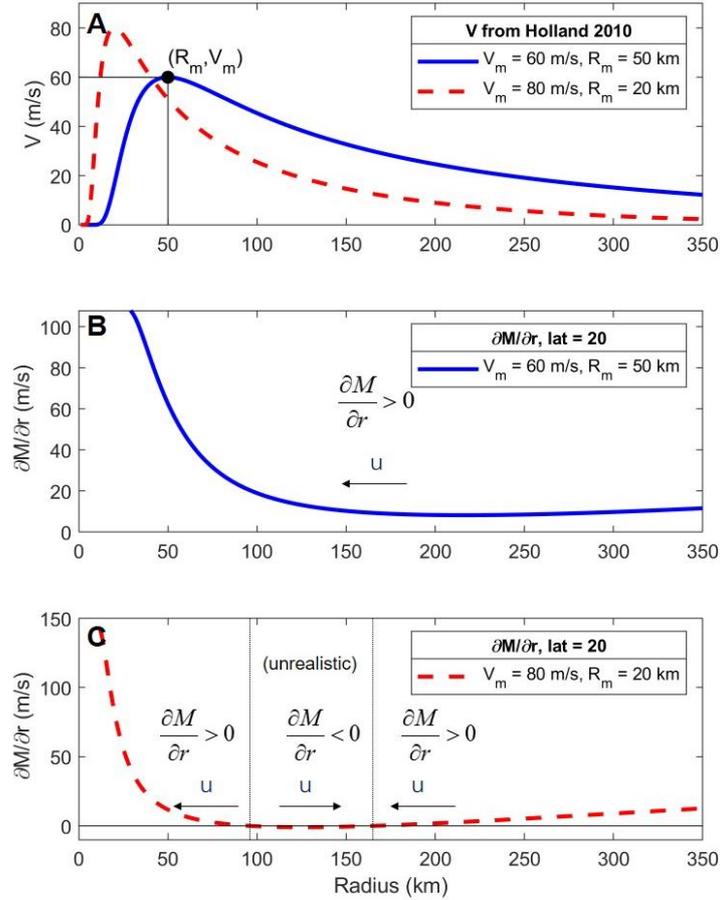

Fig. 1. Distribution of $V$ and $\partial M/\partial r$ along radius from eye for two idealized vortices. The parametric wind model used is Holland's 2010 (Holland et al. 2010).

Figure 1A shows $V$ constructed from two pairs of $V_m$ and $R_m$ with a widely used parametric wind model, Holland's 2010. For ordinary TCs (blue line in Fig. 1), with a predefined $V$ computed from observed $V_m$ and $R_m$, the radial gradient of angular momentum ($\partial M/\partial r$) remains positive along $r$, resulting a negative $u$ from Eqn. 1, indicating consistent convergence from large radius to eyewall (Fig. 1B). But for the most intense and compact TCs (featuring large $V_m$ and small $R_m$, red dashed line in Fig. 1), a predefined $V$ could make $\partial M/\partial r \to 0$ at some point along $r$ (or even negative, which will never happen in real TCs), resulting the solution of $u$ from Eqn. 1 ($u \to \infty$ or $u > 0$) unrealistic (Fig. 1C). The emergence of singularity ($\partial M/\partial r \to 0$, $u \to \infty$) demonstrates the breakdown of Eqn. 1 with such predefined $V$.

What does this singularity (breakdown) imply? The solution of $u$ from Eqn. 1 provides intuitive clues. For ordinary TCs (Fig. 1B), $u$ from Eqn. 1 is consistently negative, indicating continuous inflow from large radius to eyewall. In 'singular' cases (Fig. 1C), however, the sign of $u$ flips as $\partial M/\partial r$ goes below zero, resulting in two disconnected inflow regions along $r$ separated by an 'unrealistic' region with positive $u$. On further thought, we suspect it is a simplified picture of a storm showing secondary eyewall, which is characterized by disconnected inflow.

We next examine what pairs of $V_m$ and $R_m$, if used as inputs to construct $V$ with Holland's 2010 (Holland et al. 2010), will induce singularity in Eqn. 1. Given a certain $f$ (same as in Fig. 1), Fig. 2A shows contours of the lower bound of $\partial M/\partial r$ along $r$ with varying $V_m$ and $R_m$. We name the triangle region with $\partial M/\partial r < 0$ the 'singular zone', namely $V_m$ and $R_m$ falling in 'singular zone' induces singularity in Eqn.1. The 'singular zone' is triangle-shaped, characterized by very large $V_m$ and small $R_m$. The 'singular zone' shrinks quickly with increasing latitude (Fig. 2B), indicating a storm in higher latitudes needs to be more intense and compact to trigger secondary eyewall than in lower latitudes. This is consistent with observations that secondary eyewalls are associated with stronger wind speed, smaller eye diameter, and lower latitudes (Kossin and Sitkowski 2009; Kuo et al. 2009; Yang et al. 2013).

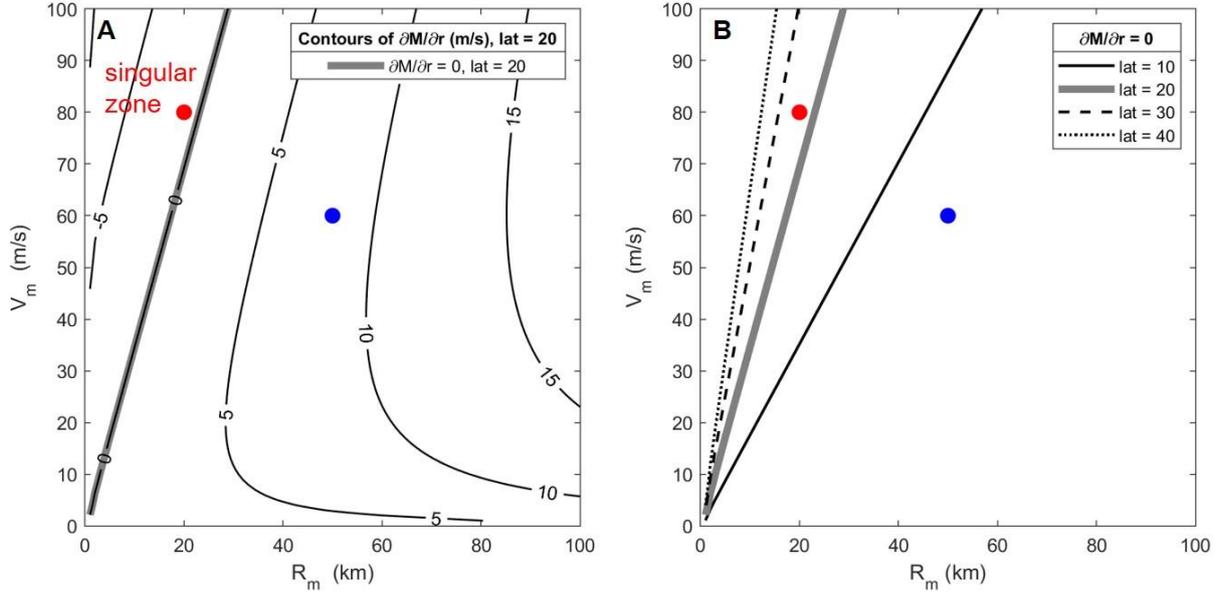

Fig. 2. (A) Contours of the lower bound of $\partial M/\partial r$ along radius for varying $V_m$ and $R_m$. (B) Contours of $\partial M/\partial r = 0$ for varying latitudes. Red and blue circles mark the two pairs of $V_m$ and $R_m$ in Fig. 1. The parametric wind model used is Holland's 2010 (Holland et al. 2010).

## 3. Evaluation

The diagnostic $\partial M/\partial r$ is dependent on three parameters: $V_m$, $R_m$, and $f$ (computed from latitude of storm center). We obtain observations of these three parameters from International Best Track Archive for Climate Stewardship (IBTrACS, Knapp et al. 2018), version 4.2. Note that while $V_m$ and $f$ are available for all storms during their lifetime, $R_m$ is only available for some storms during part of their lifetime in IBTrACS.

*a. Statistical Evaluation*

One common practice in systematically detecting secondary eyewalls is analyzing the 85 GHz channel of high resolution passive microwave data. If two rings of intense convection separated by a nearly echo-free annulus is observed, the storm is labeled as showing secondary eyewall (criteria vary slightly among studies, e.g. outer ring covers at least 2/3 of a circle (Kuo et al. 2009), or 3/4 of a circle (Kossin and Sitkowski 2009)). In this study, the secondary eyewall cases detected from passive microwave data is provided by (Kuo et al. 2009). Special Sensor Microwave/Imager (SSM/I) and the Tropical Rainfall Measuring

Mission Microwave Imager (TMI) 85 GHz data was analyzed for storms in North Western Pacific during 1997-2006. 55 out of the 225 storms were identified as secondary eyewall storms (showing an outer ring of brightness temperature $T_b \leq 230\ K$ covering at least 2/3 of a circle), which are hereafter described as the 'SE group', the rest of storms are described as the 'non-SE group'.

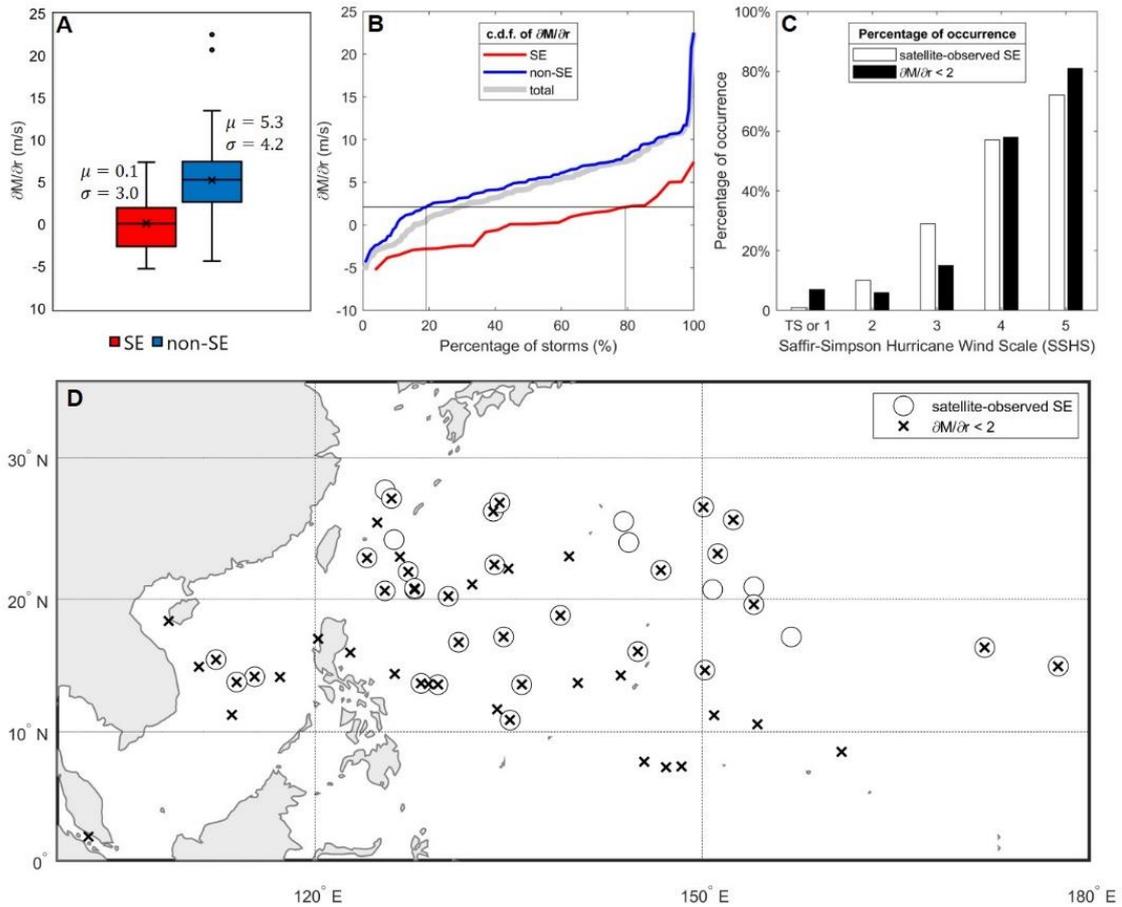

Fig. 3. Statistical evaluation of $\partial M/\partial r \to 0$ vs. satellite observations for storms in North Western Pacific 1997-2006. (A) Boxplot of $\partial M/\partial r$ for satellite-observed SE vs. non-SE storms. (B) Cumulative distribution function of $\partial M/\partial r$ for satellite-observed SE and non-SE storms. (C) Percentage of occurrence of satellite-observed SE vs. $\partial M/\partial r < 2\ \text{m s}^{-1}$. (D) Spatial distribution of satellite-observed SE cases vs. $\partial M/\partial r < 2\ \text{m s}^{-1}$ for storms with available $R_m$ from IBTrACS. Note in (D) satellite-observed SE is shown in terms of cases (36 cases) instead of storms (29 storms).

Values of $\partial M/\partial r$, lower bound along $r$ and minimum during storm lifetime for each storm, show significant differences between the 'SE group' and 'non-SE group' (Fig. 3A), with mean of 0.1 vs 5.3. This demonstrates $\partial M/\partial r$ as a specific characteristic associated with eyewall replacement. A threshold of $\partial M/\partial r = 2$ m s$^{-1}$ works best to distinguish between the 'SE group' and 'non-SE group' (able to tell about 80% of storms in both groups, Fig. 3B), and is therefore used as the practical threshold (as opposed to $\partial M/\partial r \to 0$) in the following evaluation.

A direct comparison of satellite-observed SE vs $\partial M/\partial r < 2$ m s$^{-1}$ is shown in Figs. 3C and 3D. The percentage of occurrence for $\partial M/\partial r < 2$ m s$^{-1}$ increases quickly with storm intensity, in a pattern very similar to satellite-observed SE (Fig 3C). Storms that satisfy the two criteria show similar spatial distributions too, with most of the cases distribute between 10~30 degrees N (Fig 3D).

The number of storms recognized by $\partial M/\partial r < 2$ m s$^{-1}$ is higher than satellite-observation (Fig 3D). One of the reasons may be satellite-observation is scattered in time, there are chances that SE occurs with no satellites passing by. Another reason is $\partial M/\partial r < 2$ m s$^{-1}$ captures weak or asymmetric secondary eyewall structures that do not satisfy $T_b \leq 230$ $K$ covering 2/3 of a circle, and therefore not recognized as SE by (Kuo et al. 2009). One notable example is storm Vamei 2001 close to Equator (102 E, 2 N in Fig. 3D). It is among the ~20% (Fig. 3B) of 'non-SE' storms with $\partial M/\partial r < 2$ m s$^{-1}$. Vamei is a weak storm with peak intensity of only 33 m s$^{-1}$ (SSHS category 1), but associated with $\partial M/\partial r < 2$ m s$^{-1}$ for over 48 hours (Appendix A. fig. 1). Microwave images of Vamei do show a secondary ring of intense convection, although both the inner and outer rings highly asymmetric (Appendix A. fig. 1). Like Vamei, storms with $\partial M/\partial r < 2$ m s$^{-1}$ but *not* recognized as SE by satellite-observation, due to weak intensities or asymmetries, worth further investigation in the future.

*b. Case study evaluation: timing*

The eyewall replacement of Hurricane Rita (2005) has been extensively observed and analyzed in the literature (e.g. Houze et al. 2007; Bell et al. 2012) and is used here as a case study. Fig. 4 shows time serial of $\partial M/\partial r$ for Rita, lower bound along $r$ during storm lifetime computed from $V_m$, $R_m$, and $f$ observations documented in IBTrACS. The times of observed SE occurrence reported by previous literature (marked by black circles), fall right into the time window with $\partial M/\partial r < 2$ m s$^{-1}$ (shading). Similarly, time series of $\partial M/\partial r$ and reported times of observed SE are compared for 12 frequently-discussed storms in literature (Fig. 5). Most observed SE occurrences fall in, or very close to, the time window with $\partial M/\partial r < 2$ m s$^{-1}$. This demonstrates the skill of the diagnostic to roughly capture the timing of SE occurrence.

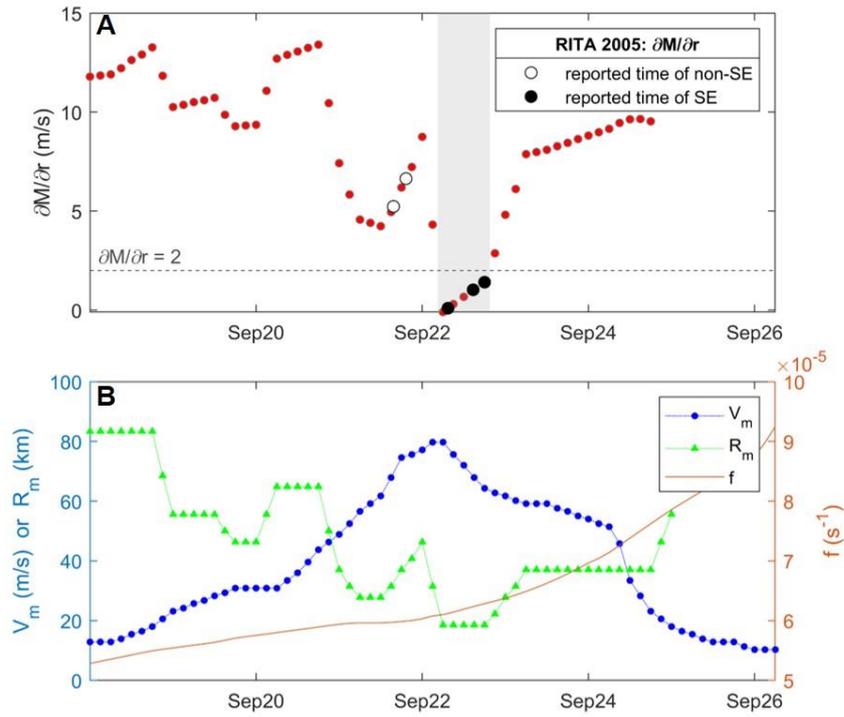

Fig. 4. (A) Time serial of $\partial M/\partial r$ and reported time of SE / non-SE in Houze et al. (2007) and Bell et al. (2012) for Hurricane Rita. (B) $V_m$, $R_m$, and $f$ for Hurricane Rita from IBTrACS. The shading in (A) indicates the time window associated with $\partial M/\partial r < 2$ m s$^{-1}$.

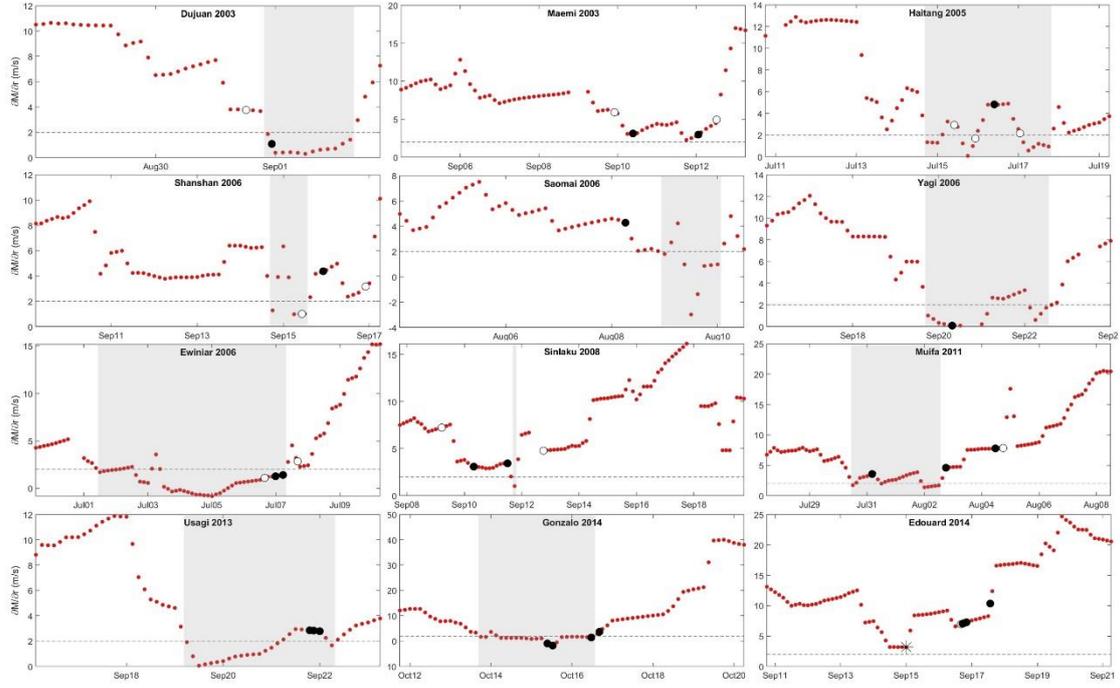

Fig. 5. Time serials of $\partial M/\partial r$ for 12 secondary eyewall storms that are frequently discussed in literatures. Shading indicates the time window associated with $\partial M/\partial r < 2$ m s$^{-1}$. Black/white circles mark reported time of observed SE / non-SE. Star (only one in Hurricane Edouard 2014) indicates a secondary eyewall event that is found by authors of this study (Appendix A. fig. 2) but not reported in literatures.

Although the value of $\partial M/\partial r$ is a combined result of $V_m$, $R_m$, and $f$, it's clear from Fig. 4 that the temporal variation of $\partial M/\partial r$ is largely shaped by the temporal variation of $R_m$. In most cases, $R_m$ documented in IBTrACS is radius of maximum wind of the *inner* vortex. But during eyewall replacement, especially when outer eyewall dominating the inner one, $R_m$ documented in IBTrACS may be the radius of the *outer* vortex, and if used as input to the diagnostic, may cause $\partial M/\partial r$ bias towards larger values. This might explain why some observed SE are associated with relatively large values of $\partial M/\partial r$, immediately following the window of $\partial M/\partial r$ being small (e.g. Shanshan and Muifa in Fig. 5).

It's worth noting that some observed SE are associated with relatively large values of $\partial M/\partial r$ being stable for over 18 hours, e.g. Muifa after Aug 4, and Edouard after Sep 16 (Fig. 5). These are typical examples of satellite-observed SE cases associated with $\partial M/\partial r > 2$ m s$^{-1}$. According to Fig. 3B, such storms account for ~20% of satellite-observed SE storms. It is a clear sign that there exist multiple mechanisms for SE occurrence, and some do not directly involve $V_m$, $R_m$, $f$, such as the interaction of the vortex with mid-latitude jet or upper-level trough. The attribution of different SE mechanisms is a very interesting topic for future studies.

## 4. Discussions

*a. Singularity as a diagnostic*

The eyes of TCs have long been viewed as singularities of the atmospheric systems, although the explicit mathematical expression of such singularity has not yet been revealed. If we shift our focus from intense convection of the storm, i.e. eyewalls, to the opposite of them, i.e. the eye and the moat, then eyewall replacement is not only the emergence of an outer eyewall gradually contracts and replaces inner eyewall, but also the emergence of a moat that gradually joins the eye.

In this study, the occurrence of secondary eyewall, or the occurrence of the moat, is found to take form of a singularity ($\partial M/\partial r \to 0$). Given that the moat is "dynamically similar to the eye" (Houze et al. 2007) and eventually becomes part of the eye, we are thrilled to think this diagnostic may shed light on the explicit form of singularity of the eye.

*b. Correspondence with literature*

This diagnostic depicts eyewall replacement as a storm entering the 'singular zone' on $V_m - R_m$ plane (Fig. 2), with the 'singular zone' shrinks quickly with increasing latitudes. This corresponds nicely with observations that secondary eyewalls are associated with stronger wind speed, smaller eye diameter, and lower latitudes (Kossin and Sitkowski 2009; Kuo et al. 2009; Yang et al. 2013).

Environmental conditions (e.g. higher sea surface temperature, weaker environmental wind shear (Kossin and Sitkowski 2009; Yang et al. 2013), higher humidity(Hill and Lackmann 2009)) and internal processes (e.g. the wind-induced surface heat exchange (Nong and Emanuel 2003; Cheng and Wu 2018), outer-core latent heating (Bell et al. 2012; Rozoff et al. 2012)) that favor the development of intense and compact storms will contribute to the development of secondary eyewall by pushing the storm into the 'singular zone'. But once inside the 'singular zone' and eyewall replacement starts (moat forms), the intensity of inner eyewall will no longer respond to the favoring environment. This explains why the drop of $V_m$ during eyewall replacement typically occurs in an environment that is not indicative of weakening (Kossin and DeMaria 2016), and accompanied with maintaining or increasing convective activity (Yang et al. 2013).

This diagnostic corresponds nicely with existing eyewall replacement theories as well. The emergence of this singularity in computed inflow in TC boundary layer is a direct representation of unbalanced boundary layer processes, as emphasized in Bell et al. (2012), Huang et al. (2012), Abarca and Montgomery (2013, 2014), Kepert (2013). The fact that Holland's 2010 wind model (Holland et al. 2010) works well in showing the singularity, might be because *gradient wind balance* is assumed in Holland's, which cannot accommodate the emergence of *supergradient wind*, the key feature in unbalanced TC boundary layer. This diagnostic also gives a clear sign that there exist multiple mechanisms for secondary eyewall occurrence, and some do not directly involve $V_m$, $R_m$ and $f$, such as the interaction of the vortex with mid-latitude jet (Dai et al. 2017) or upper-level trough (Nong and Emanuel 2003; Molinari and Vollaro 1990).

## 5. Summary

We propose a simple diagnostic for secondary eyewall occurrence, $\partial M/\partial r \to 0$, i.e. the storm entering the 'singular zone' on $V_m - R_m$ plane. This diagnostic is a direct representation of unbalanced boundary layer processes and the emergence of supergradient wind, which is shown to account for ~80% satellite-observed secondary eyewall storms.

Despite its great impact on storm size and intensity, secondary eyewall has rarely been accounted for in statistical intensity prediction or risk assessment of TCs. The simple form of this diagnostic makes it possible. We look forward to improvements in the modeling of TC intensity and TC-induced hazards, especially TC rainfall modeling, where the singularity was first observed.

*Acknowledgments.*

We are truly grateful for the shared data from Prof. Hung-Chi Kuo, comments and encouragement from Prof. Kerry Emanuel, and detailed feedback and suggestions from Rohini Shivamoggi. This study is based upon work supported by the National Natural Science Foundation of China grant 42005116 (PL) and grant 42130603 (YL).

*Data Availability Statement.*

Information of satellite-observed SE and non-SE are included in Kuo et al. (2009).

IBTrACS dataset is openly available from NOAA NCEI at https://doi.org/10.25921/82ty-9e16 as cited in Knapp et al. (2018, 2010).

SSM/I dataset is openly available from NOAA NCEI at https://doi.org/10.7289/V5SJ1HKZ as cited in Wentz et al. (2013).

TMI dataset is openly available from NASA GES DISC at https://doi.org/10.5067/GPM/TMI/TRMM/1B/05 as cited in TMI (2017).

# APPENDIX

## Appendix A. Vamei and Edouard

Time serials of $\partial M/\partial r$ and microwave images for Storm Vamei (2001) and Hurricane Edouard (2014).

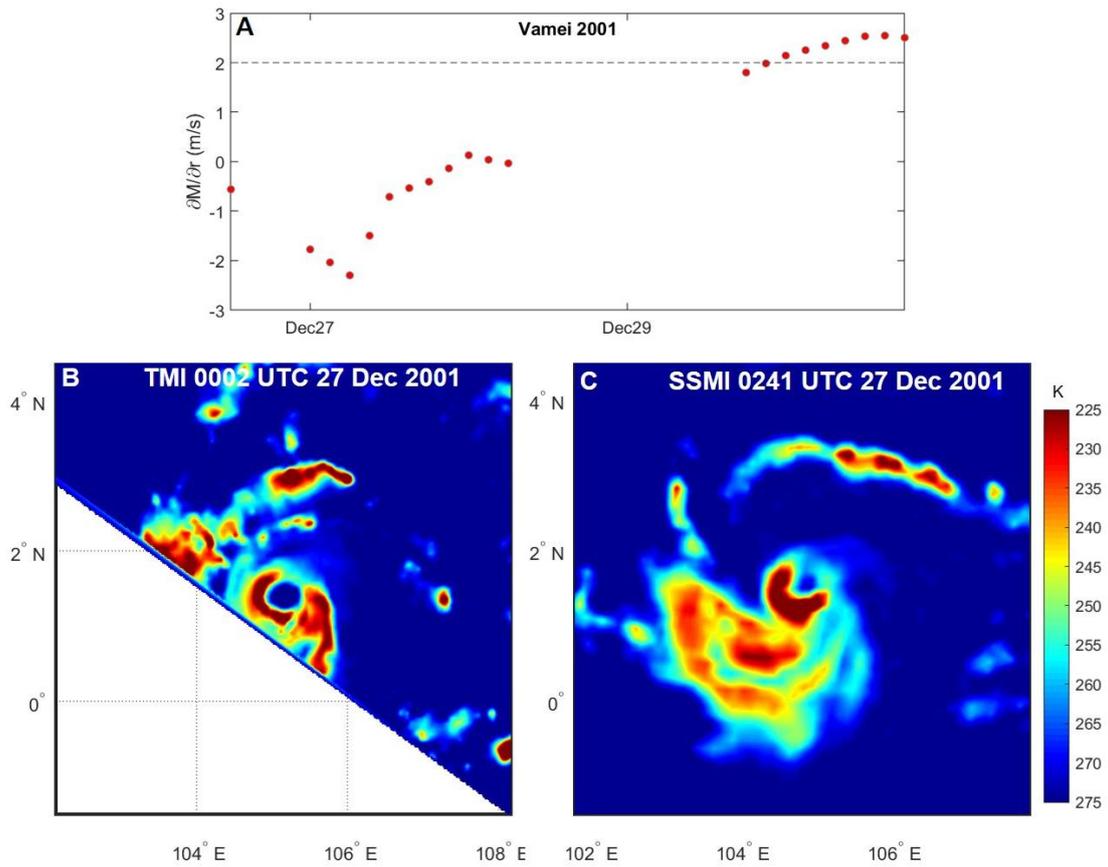

fig. 1. (A) $\partial M/\partial r$ computed from IBTrACS shows $\partial M/\partial r < 2$ m s$^{-1}$ for over 48 h. (B,C) Microwave images from TMI (Precipitation Processing System (PPS) At NASA GSFC 2017a) and SSM/I (Wentz et al. 2013) showing the storm with an outer ring of intense convection, and the storm (the primary vortex, the moat, and the outer ring) being highly asymmetric.

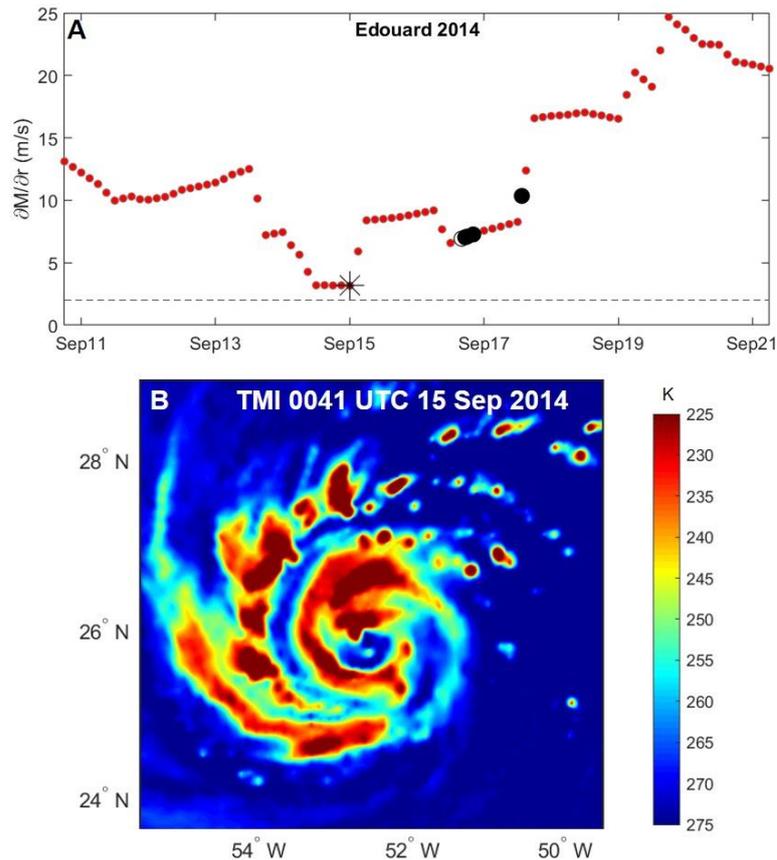

fig. 2. (A) $\partial M/\partial r$ computed from IBTrACS. Black/white circles mark reported time of observed SE / non-SE, star indicates a secondary eyewall event not reported in literatures. (B) Microwave image from TMI (Precipitation Processing System (PPS) At NASA GSFC 2017a) at 0041 UTC 15 Sep 2014 shows an outer ring of intense convection, and the storm (both inner and outer rings) being asymmetric.